\documentclass{ws-procs9x6}
\def\ltsim{\hbox{\raise 2pt \hbox {$<$} \kern-1.1em \lower 4pt \hbox {$\sim$}}}
\def\ltapprox{\hbox{\raise 2pt \hbox {$<$} \kern-1.1em \lower 5pt \hbox 
{$\approx$}}}
\def\gtsim{\hbox{\raise 2pt \hbox {$>$} \kern-1.1em \lower 4pt \hbox {$\sim$}}}
\def\gtapprox{\hbox{\raise 2pt \hbox {$>$} \kern-1.1em \lower 5pt \hbox 
{$\approx$}}}

\def\arcmin{$^{\prime}$}

\begin{document}

\title{Non-thermal phenomena in galaxy clusters}

\author{Luigina FERETTI }

\address{Istituto di Radioastronomia CNR \\
Via P. Gobetti 101 \\ 
40129 Bologna, Italy\\ 
E-mail: lferetti@ira.cnr.it}


\maketitle

\abstracts{
The recent evidences of the presence of magnetic fields and
relativistic particles in galaxy clusters are reviewed.
The existence of $\mu$G-level magnetic fields in cluster atmospheres
appears well established from the detection of diffuse radio emission
and from studies of Rotation Measure. 
The fact that the diffuse radio emission is not common in clusters
favours the hypothesis that the population of relativistic particles,
produced during the cluster formation by AGNs or star formation, are
reaccelerated by recent cluster merger processes.
}

\section{Introduction}

It is nowadays well established that 
the clusters of galaxies contain two relevant
non-thermal components: magnetic fields and
relativistic particles.

The most detailed evidence for these components
comes from the radio observations.
A number of clusters of galaxies is known to contain large-scale
diffuse radio sources (radio halos and relics) 
which have no obvious connection with the
cluster galaxies, but are rather associated with the
intracluster medium (ICM). The synchrotron origin of the emission
from these sources requires a population of 
GeV relativistic electrons and cluster-wide
$\mu$G magnetic fields in the ICM.

Indirect evidence of  the existence of cluster magnetic fields
derives from studies of the Rotation Measure of
radio galaxies embedded within the cluster thermal atmospheres
or located behind them.
Probe of the existence of a population
of relativistic particles is obtained from
the detection of non-thermal emission of inverse Compton origin
in the hard X-ray (HXR) and extreme ultraviolet (EUV) domains.

The instrinsic parameters quoted in this paper are computed
with a Hubble constant H$_0$ = 50 km s$^{-1}$ Mpc$^{-1}$ 
and a deceleration parameter q$_0$ = 0.5.

\section{Radio halos, relics and mini-halos}

Great attention has been devoted in recent years to the study of
cluster large-scale diffuse radio sources.  New halo and relic
candidates were found from searches in the NRAO VLA Sky 
Survey\cite{gtf99}, in the Westerbork Northern Sky
Survey\cite{ks01} and in the survey of the Shapley 
Concentration\cite{ven00}. 
The number of clusters with halos and relics is
presently around 50. Their properties have been recently reviewed by
Giovannini and Feretti\cite{gfbo}.

Radio {\it halos} are diffuse radio sources of low surface
brightness ($\sim$ $\mu$Jy arcsec$^{-2}$ at 1.4 GHz) and
approximately regular shape, similar to the diffuse source
Coma C at the center of the Coma cluster, 
first classified by Willson\cite{w70}.
They are typically extended \gtsim~1 Mpc (although smaller halos down to 
\ltsim~500 kpc in size have also been detected), are unpolarized  
and show  a steep
radio spectrum typical of aged radio sources ($\alpha$ \gtsim~ 1).
The spectrum of Coma C shows a radial 
decrease\cite{gcoma} from  $\alpha \sim$ 0.8 at the cluster center, 
to $\alpha \sim$ 1.8 beyond a distance of about 10\arcmin.

Recently, radio halos have been
studied in distant clusters, as A665\cite{gf00} (z = 0.1818), 
A2163\cite{f01} (z = 0.203), A2744\cite{gov01} (z = 0.308), 
and CL 0016+16\cite{gf00} (z = 0.5545).  
A powerful radio halo was found in the hottest known
cluster of galaxies 1E$0657-56$\cite{l00} (kT = 15.6 keV; z = 0.296).
The existence of a complex halo and a
possible relic has been confirmed in A754\cite{k01}$^,$\cite{b03},
 where the presence of diffuse emission was debated in the
literature.

A possibly related phenomenon to radio halos is a class of sources
found in the cluster peripheral regions, the {\it radio relics}. 
Relic sources are similar to halos in their low surface brightness,
large size and steep spectrum.
Unlike halos, they generally show an elongated structure and are highly
polarized.  
The prototype of this class is 1253+275,
in the Coma cluster, first classified by Ballarati et al.\cite{b81}.  
A spectacular example of two almost symmetric relics in the
same cluster is found in A3667\cite{r97}.
A puzzling relic source is 0917+75\cite{har93}$^,$\cite{gf00},
located at 5 to 8 Mpc from the centers of the closest
clusters (A762, A786, A787), thus not unambiguously associated with
any of them.

Feretti\cite{f00} argued that halos and relics are not the same objects
seen in projection, i.e. halos are really at the cluster center and
not simply projected onto it.  Halos and relics may indeed have
different physical origins (see Sect. 6).

{\it Mini-halos} are diffuse radio sources of moderate size ($\sim$
500 kpc) surrounding a dominant powerful radio galaxy at the center of
cooling flow clusters, as detected in the
Perseus and Virgo clusters. From a study of the Perseus mini-halo\cite{gbs02},
it was argued that a connection between the central radio 
galaxy and the mini-halo
in terms of particle diffusion or buoyancy is not possible. 
This lead to the suggestion that the relativistic electrons are
reaccelerated by turbulence in the cooling flow region of the 
ICM. Deeper studies of a large sample of mini-halos are needed to clarify
the connection between the central radio galaxy and the formation
of mini-halos.

\subsection{Physical conditions}

The physical parameters in radio sources can
be estimated assuming a minimum energy configuration for the
summed energy in relativistic particles and magnetic fields.
This roughly corresponds to energy equipartition
between fields and particles.  
The derived minimum energy densities in halos and relics
are of the order of 10$^{-14}$ - 10$^{-13}$ 
erg cm$^{-3}$, i.e.  much lower than the energy density
in the thermal gas.
These calculations typically assume 
equal energy  in relativistic protons and electrons,
a volume filling factor of 1, a
low frequency cut-off of 10 MHz, and a 
high frequency cut-off of 10 GHz.
The corresponding equipartition
magnetic field strengths  range from  0.1 to 1 $\mu$G.

Due to synchrotron and inverse Compton losses, the typical lifetime
of the relativistic electrons in the ICM is relatively
short ($\sim$10$^8$ yr)\cite{sar99}. 
 The difficulty in explaining
radio halos and relics arises from the combination of their large size
and the short synchrotron lifetime of relativistic electrons.  The
expected diffusion velocity of the electron population is of the order
of the Alfv\'en speed ($\sim$ 100 km s$^{-1}$) making it difficult for
the electrons to diffuse over a megaparsec-scale region within their
radiative lifetime.  Thus the relativistic particles need to be
reaccelerated by some mechanism, acting with an efficiency comparable
to the energy loss processes\cite{pet01}.  We will show in the
following that recent cluster mergers are likely to supply energy to the
halos and relics.  

\subsection{Connection to cluster merger processes}

Unlike the presence of thermal X-ray emission, the presence
of diffuse radio
emission is not  common  in clusters of galaxies:
in a complete cluster sample,
5\% of clusters have a radio halo source and 6\% a peripheral relic
source\cite{gfg00}.  The detection rate of diffuse radio sources increases with
the cluster X-ray luminosity, reaching $\sim$35\% in clusters with
X-ray luminosity larger than 10$^{45}$ erg s$^{-1}$ (in the Rosat
band 0.1-2.4 keV).

To explain the relative rarity of diffuse sources, it has been
argued that the formation of halos and relics is connected to
recent cluster merger events. Indeed, mergers generate shocks, 
bulk flows, turbulence in the ICM. These processes would
provide energy to reaccelerate the radiating
particles all around the cluster.

Several evidences favour the hypothesis that clusters with halos and
relics are characterized by strong dynamical activity,
likely related to merging processes.  These clusters indeed
show: i) substructures and distortions in the X-ray
brightness distribution\cite{sch01}; ii)
 temperature gradients\cite{mar98}; iii) 
absence of a strong cooling flow\cite{fer99}; 
iv) values of spectroscopic $\beta$ on
average larger than 1\cite{f00}; v)  core radii significantly
larger  than those of clusters
classified as single/primary\cite{f00};
vi) large distance from the nearest neighbours\cite{sch02};
vii) large values of the cluster dipole power ratio\cite{buo01}. 

High resolution X-ray Chandra
data have been recently obtained for several clusters
with halos or relics, e.g. A665\cite{mv01}, 
A2163\cite{mv01}, 1E 0657-56\cite{mar02},
A520\cite{mtaiw}, A3667\cite{mtaiw}.
In all these clusters, temperature gradients and gas
shocks are detected, confirming the presence of mergers.
Preliminary maps of the radio spectral index between 0.3 and 1.4 GHz
in  A665 and A2163 (Feretti et al.  in preparation)
show possible connection to the Chandra temperature maps, indicating
a link between halos and cluster mergers.

In conclusion, there seems to be convincing evidence that diffuse
sources are preferentially associated with high X-ray luminosity
clusters with recent mergers. 
We are not presently aware of any radio halo or relic  in
a cluster where the presence of a merger has been clearly
excluded. On the other hand, not all merging clusters
host a diffuse radio source.

\subsection{Halo radio power vs cluster parameters} 

The most powerful radio halos  are detected 
in the clusters with the highest X-ray luminosity. This follows
from the  correlation found between the diffuse source
radio power and the cluster X-ray luminosity\cite{l00}$^,$\cite{b03}.
An extrapolation of the above correlation to low radio and
X-ray luminosities indicates that clusters with L$_{X}$
\ltsim~10$^{45}$ erg s$^{-1}$ would host halos of power of a few
10$^{23}$ W Hz$^{-1}$.  With a typical size of 1 Mpc, they would have
a radio surface brightness lower than current limits obtained in the
literature and in the NRAO VLA Sky Survey.  Therefore, it is possible
that future new generation instruments (LOFAR, SKA) will allow the
detection of low brightness/low power large halos in virtually all the
merging clusters.  
 
It is worth to remind that the correlation is valid
for  merging clusters with radio halos,  and
therefore cannot be generalized to all clusters.  Among the clusters
with high X-ray luminosity and no radio halo, there are A478, A576,
A2204, A1795, A2029, all well known relaxed clusters with a massive
cooling flow. 

Since cluster X-ray luminosity and mass are correlated\cite{rb02}, 
the  correlation between radio power and X-ray
luminosity could reflect a dependence of the  radio power on the
cluster mass. This is indeed the case: 
a correlation of the type P$_{\rm 1.4~GHz}$ $\propto$ M$^{2.3}$ is 
derived for radio halos\cite{gov01}$^,$\cite{ftaiw}. This is
similar to what expected from simple theoretical considerations.
Actually, assuming that roughly the energy released in a merger shock
is proportional to the gas density $\rho$ and to the third power 
of the subcluster
velocity $v^3$, and that $\rho \propto$ M, and $v\propto {\rm
M}^{1/2}$, it is obtained that \.E $\propto$ M$^{5/2}$.  
This could indicate that the cluster mass may be
a crucial parameter in the formation of radio halos\cite{buo01}.  Since it
is likely that massive clusters are the result
of several major mergers, we conclude that 
both past mergers and current mergers
are the necessary ingredients for the formation and evolution of radio
halos. This scenario may provide a further explanation
of the fact that not all clusters showing recent mergers host radio halos.

\section {Evidence of cluster magnetic fields}

Synchrotron radiation from cosmic radio sources is well known to be
linearly polarized.
A linearly polarized wave of wavelength $\lambda$,  
traveling from a radio source through a magnetized medium, 
experiences a phase shift of the left versus right circularly
polarized components of the wavefront, leading to a
rotation $\Delta\chi$ of the position angle of the polarization,
according to the law: $\Delta\chi$ = RM $\lambda^2$,
where RM is the Faraday rotation measure.
The RM is related to the electron density, $n_{\rm e}$
(in 10$^{-3}$ cm$^{-3}$), and to the magnetic field, $\bf B$
(in $\mu$G), as:
$${\rm RM} = 0.812\int\limits_0^L n_{\rm e} {\bf B} \cdot d{\bf l}
~~~~~~~~~~{\rm  radians~m}^{-2}~$$
where the path length {\bf l} is in kpc.
The RM values can be derived from multifrequency
polarimetric observations of sources within or behind the
clusters, by measuring the
position angle of the polarized radiation as a
function of frequency.
They can then be combined with measurements of
$n_{\rm e}$ to estimate the cluster magnetic field along the line
of sight.

This kind of studies has been performed on several individual
clusters, with or without cooling flows, and
on statistical samples (see the review by Carilli and 
Taylor\cite{ct02} and references therein).
In general, the suggestion from the data is 
that  magnetic fields in the range of 1-5 $\mu$G are common
in clusters, regardless of the presence or not of
diffuse radio emission. At the center of cooling flow clusters,
magnetic field strenghts can be larger by about a factor of 2.
Another result of the above mentioned studies
is that the RM distributions tend to be patchy with
coherence lenghts of 5-10 kpc, indicating
that the magnetic fields are not ordered on cluster (Mpc)
scales, but consist of cells with random field orientation.

Thus, in most clusters the fields are not dynamically important,
with magnetic pressures much lower than the thermal pressures,
but the fields may play a fundamental role in the energy
budget of the ICM.
The simplest model is a uniform field throughout the cluster. However,
possible filaments and flux ropes may be present\cite{eil99}. 
Also, the value of the magnetic field intensity is likely to decrease
with the distance from the cluster center, as in Coma\cite{brcom} 
and in A119\cite{dol01} (here 
the field scales as $n_e^{0.9}$). 

The ICM magnetic fields could be primordial,
or injected  from galactic winds, or from active 
galaxies, or produced in shock waves of the large scale
structure formation.
The seed fields then need to be amplified to give the fields that we
observe at present. 
The most likely possibility is that the magnetic fields are 
amplified by turbulence following a
cluster merger. Simulations\cite{rot99}
show that the magnetic field energy increases by greater than a factor
10-20 in localized regions during mergers. 
It is likely that massive clusters undergo 
several major mergers during their lifetime
and that each successive merger will further amplify the fields.
Numerical studies\cite{dol00}$^,$\cite{rot99}  
 of hierarchical merging of large scale structure
including an initial intergalactic field of $\sim 10^{-9}$ G
show that a combination of
adiabatic compression and non-linear amplification in shocks
during  cluster mergers may lead to ICM mean fields of the
 order of $\sim \mu$G.

\section{Relativistic particles from Inverse Compton emission}

The relativistic electrons present in the ICM produce
inverse Compton (IC) radiation because of scattering with the
cosmic background photons. Depending on the electron Lorentz
factor $\gamma$, this radiation is detected as
hard X-ray emission ($\gamma \sim$ 10$^{4}$) or
extreme ultraviolet emission ($\gamma \sim$ 3 10$^{2}$).

\subsection{Hard X-ray Emission (HXR)} 

The high energy relativistic electrons, with $\gamma$ $\sim$ 10$^4$, 
responsible for the radio emission in the ICM, 
scatter off the cosmic microwave
background (CMB), boosting photons from this radiation field
to the X-ray and $\gamma$-ray regions. 
Measurements of this radiation provide additional information that, when
combined with results of radio measurements
(i.e. the ratio of hard X-ray IC emission to radio synchrotron 
emission), enables the
determination of the electron density and mean  magnetic field
directly, without the need to invoke equipartition.
If the X-ray and radio emissions are produced by the same population of 
electrons undergoing inverse Compton and synchrotron energy losses, 
respectively, the main 
expected features of the X-ray emission are: a) power law spectrum with an 
index which is  related to the radio index; b) X-ray to radio 
luminosity ratio roughly equal to the ratio between
the CMB energy density
and the magnetic field energy density.

A significant breakthrough in the measurement
of HXR emission was recently obtained owing to
the improved sensitivity and wide spectral capabilities
of  the BeppoSAX and the Rossi X-ray Timing Explorer (RXTE)  satellites.
Non-thermal hard X-ray emission at energies 
\gtsim~ 20 keV has been detected in the Coma cluster 
\cite{ff99}$^,$\cite{rep99} and in A2256\cite{ff00}.
The 20-80 keV flux in Coma is $\sim$2 10$^{-11}$ erg cm$^{-2}$ s$^{-1}$,
which leads to a magnetic field of 0.16 $\mu$G.
In A2256, the flux in the same energy range is $\sim$9 10$^{-12}$
erg cm$^{-2}$ s$^{-1}$. A magnetic field of $\sim$ 0.05 $\mu$G is
derived for the the northern cluster region, where the radio relic is detected, while a higher field
value, $\sim$0.5 $\mu$G, could be present at the cluster
center,  in the region of the radio halo.

A marginal detection has ben obtaineed in A2199\cite{kaa99},
whereas for the clusters A3667, A119, A2163 and
 A754, only upper limits to the non-thermal
X-ray emission have been derived\cite{f01}$^,$\cite{ff02}.
A possible detection of localized IC emission associated with the
radio relic and with merger shocks has been claimed in A85\cite{bag98}.

It is worth mentionining here that alternative models have been
suggested for the explanation of the hard X-ray tails 
(e.g. non-thermal bremsstrahlung\cite{bla00}$^,$\cite{dog00}$^,$\cite{sk00}).
These were motivated by the discrepancy between 
the value of the ICM magnetic field derived by the IC model
and the value
derived from Faraday rotation of polarized radiation (see Sect. 5). 
However, these models may have serious difficulties as they 
would require an unrealistic high energy input\cite{pet01}.

\subsection{Extreme Ultraviolet Emission (EUV)}

The Extreme Ultraviolet Explorer have revealed EUV (0.1 to 0.4 keV) 
emission in excess of the expected thermal emission in 
Virgo\cite{lv96}$^,$\cite{ber00} and Coma\cite{lc96}$^,$\cite{bow99}.  
The EUV detections in other clusters
(Abell 1795, Abell 2199, Abell 4038, Abell 4059 and Fornax)
 remain quite  controversial.
The EUV emission has luminosities of $\sim$10$^{44}$ erg s$^{-1}$
and has spectra which decline rapidly going from the EUV to the X-ray band.

The EUV excess may be interpreted to be of thermal origin, due to a 
relatively cool ($\sim$ 10$^6$ K) emitting gas. However,
at these temperatures gas cooling is particularly efficient, and one would
expect the presence of resonance lines which are not detected.
Therefore, a non thermal interpretation is favoured,
i.e. this emission is more likely due to IC scattering of CMB photons by
relativistic electrons. This scenario requires an electron population
with  energies of $\sim$ 150 MeV ($\gamma$ $\sim$ 300).
These electrons have a too low energy to produce detectable radio
emission. Sarazin\cite{sar99} pointed out that these electrons have lifetimes
comparable to the Hubble time, and should be present in essentially
all clusters. 
In the Coma cluster,  Brunetti et al.\cite{breuv}
suggested that the electron population injected in the central 
part of the cluster 
by the head-tail radio galaxy NGC4869 may account for a large fraction,
if not all, of the  detected EUV excess.

\section{Reconciling magnetic field derived values}

The IC estimated cluster magnetic fields
are typically 0.2 to 1 $\mu$G (Sect. 4.1). These are consistent with the
values obtained from equipartition arguments in radio halos
(Sect. 2.1). The fields derived using RM observations are instead
an order of magnitude higher (Sect. 3).

Golsdmith \& Rephaeli\cite{gr93}
suggested that the IC estimate is typically expected to be lower than the
Faraday rotation one, because of the expected spatial profiles
of the magnetic field and gas density. More recently, it has been shown 
that IC models which include the effects
of more realistic electron spectra, combined with the expected
radial profile of the magnetic field, and anysotropies
in the pitch angle distribution of the electrons allow higher values
of the ICM magnetic field in better agreement with the Faraday rotation
measurements\cite{brcom}$^,$\cite{pet01}.

For example, if the magnetic field strength has a radial decrease,
most of the IC emission will come from the weak field
regions in the outer parts of the cluster, while most of the Faraday
rotation and synchrotron emission
occurs in the strong field regions in the inner parts of the
cluster.

Recent modeling by Govoni \& Murgia (in preparation)
shows that the magnetic field substructure and/or filamentation
can lead to significant differences between field estimates obtained
from different approaches. 

Finally, it has been recently pointed out that in some cases a radio
source could compress 
the gas and fields in the ICM to produce local RM 
enhancements, thus leading to overestimates of the derived ICM 
magnetic field strenght\cite{rud02}.

\section{Models for Relativistic Particles}

A population of relativistic electrons can account 
for the radio halos and the HXR and EUV emission in clusters
via synchrotron  and inverse Compton
processes, respectively. Current models have been reviewed
by Brunetti\cite{brtaw}. The relativistic particles could be injected
in the cluster volume from AGN activity 
(quasars, radio galaxies, etc.), 
or from star formation in normal galaxies
(supernovae, galactic winds, etc).
Most of the particle production has occurred in the past
and is therefore connected to the dynamical history of the clusters.

This population of {\it primary electrons}  needs to be 
reaccelerated\cite{brcom}$^,$\cite{pet01} to compensate the radiative losses. 
The hypothesis  that a recent cluster merger is the most likely process
acting in  the reacceleration of relativistic particles has been 
worked out in recent years. 
In major mergers, hydrodynamical shocks dissipate energies of 
$\sim$ 3 10$^{63}$ erg, which could be 
partly converted into the acceleration of
relativistic electrons\cite{sar99}$^,$\cite{tn00}$^,$\cite{min01}.
The shock acceleration has been recently questioned\cite{gb03}. 
Alternatively, stochastic acceleration by a turbulent ICM 
seems to be an efficient process\cite{brcom}$^,$\cite{fts03}.
The most likely scenario appears to be an
episodic injection-acceleration
model, whereby one obtains a time dependent spectrum
that for certain phases of its evolution satisfies all
the requirements\cite{pet01}. 

Another class of models for the radiating particles in halos
involves {\it secondary electrons}, resulting from inelastic 
nuclear collisions between the relativistic protons and the 
thermal ions of the ambient intracluster medium. 
The protons diffuse on large
scales because their energy losses are negligible. 
They  can continuously produce in situ electrons, distributed 
through the cluster volume\cite{bc99}$^,$\cite{min01}.
In the framework of secondary electron models, it is difficult
to explain the observed association between mergers and radio halos,
the  spectral index radial steepening found in Coma C and the
relatively low number of clusters with halos.
Strong $\gamma$-ray emission, which should be detected by future
$\gamma$-ray instruments,  
is expected to be produced in this case, providing 
a test of the secondary electron models.

Different models have been suggested for the origin 
of the relativistic electrons radiating in the radio relics, i.e.
located in confined peripheral regions of the clusters.
There is increasing evidence
that the relics are tracers of shock waves in merger events\cite{e98}.
Active radio galaxies may fill
large volumes in the ICM with radio plasma, which becomes rapidly 
invisible to radio telescopes because of radiation losses of the
relativistic electrons.  These patches of fossil radio plasma
are revived by adiabatic 
compression in a shock wave
produced in the ICM by the flows of cosmological large-scale
structure formation\cite{e99}$^,$\cite{egk01}.

\section{Conclusions}

Diffuse radio emission is detected in X-ray luminous, massive clusters
showing strong dynamical activity and merger processes. It 
demonstrates the existence of magnetic fields and relativistic particles
in the ICM.

Studies of rotation measure of radio sources within or behind
clusters indicate the presence of large scale magnetic fields of
the order of $\sim \mu$G in the majority of clusters, not only
in clusters with halos. These fields are likely to result from
the amplification of seed fields during the mergers occurring
in the cluster formation process.

The population of relativistic particles produce
inverse Compton radiation in HXR and EUV. This emission has been presently
detected only in a few clusters. Since the 
radiating particles have short lifetimes, they need to be 
continuously reaccelerated. 

There is a general consensus
that recent merger phenomena would provide the energy for the relativistic
electron reacceleration, thus allowing the production of  
a detectable diffuse radio emission.
This emission is also likely related to the cluster dynamical 
history.

\section*{Acknowledgments}

I would like to thank the organizers for making possible this
enjoyable and scientifically profitable conference.
This work was partially funded by the Italian Space Agency.

\end{document}